\documentclass{elsart5p}
\usepackage{ifpdf}
\usepackage{graphicx,amssymb,lineno,subfigure}
 \usepackage{graphics}
 \usepackage{epsfig}

\usepackage{amssymb}

\begin{document}

\begin{frontmatter}



\title{The Nonlinear Dirac Equation in Bose-Einstein Condensates: \\Foundation and Symmetries}


 \author{L. H. Haddad and}
 \author{L. D. Carr}
 \address{ Department of Physics, Colorado School of Mines, Golden, Colorado 80401, USA}
 \ead{lhaddad@mines.edu}
\ead{lcarr@mines.edu}

\begin{abstract}
We show that Bose-Einstein condensates in a honeycomb optical lattice are described by a nonlinear Dirac equation in the long wavelength, mean field limit.
Unlike nonlinear Dirac equations posited by particle theorists, which are
designed to preserve the principle of relativity, i.e., Poincar\'e covariance, the nonlinear Dirac equation for Bose-Einstein condensates breaks
this symmetry.  We present a rigorous derivation of the nonlinear Dirac equation from first principles.  We provide
a thorough discussion of all symmetries broken and maintained.
\end{abstract}

\end{frontmatter}

% main text
\section{ Introduction}
\label{sec:1}
Recently the first truly two-dimensional (2D) solid state material, graphene, was created in the laboratory~\cite{novoselov2005,zhangYB2005}.  One of the most exciting aspects of this novel material is that
long wavelength excitations are described by a Dirac equation for massless particles,
with a ``speed of light'' equal to the
Fermi velocity $v_F \simeq c/300$~\cite{geim2007}.
Thus one can study relativistic phenomena at very low velocities
in an experiment far more accessible than a particle accelerator.
The only real requirement to obtain this equation is the simple hexagonal, or honeycomb lattice
structure of the graphene~\cite{wuC2007}.  One can therefore consider any solid state system constructed
on a honeycomb lattice, including artificial systems, in order to study relativistic phenomena in
novel materials accessible in tabletop experiments.

The most precise, cleanest, most controllable artificial solid state system is
ultra-cold atoms in optical lattices.  Such systems have no impurities and no disorder
unless specifically added in by hand.  They are very versatile: they can be constructed with
an arbitrary lattice structure in one, two or three dimensions; they can contain bosons and/or fermions,
atoms and/or diatomic molecules; and they can even have a pseudo-spin structure.
Their temperature and interaction sign, magnitude, and symmetry can be controlled externally.
Moreover, 2D physics has recently been of great interest
in this context, in the form of the Berzinskii-Kosterlitz-Thouless crossover~\cite{hadzibabic2006},
and 2D systems underpinned by lattices are immediately available in experiments.
Instead of considering ultra-cold fermions, which could be used to produce an near-exact
analog of graphene~\cite{zhuSL2007}, we consider ultra-cold bosons.  Bosonic statistics
and interactions lead to a new feature in the massless Dirac equation known in graphene -- a naturally occurring nonlinear term, giving rise to a \emph{nonlinear Dirac equation}.

The study of nonlinear phenomena in ultra-cold atoms, especially in Bose-Einstein condensates (BECs)~\cite{dalfovo1999,leggett2001}, has been enormously fruitful.
The recent text edited by Kevrekidis, Frantzeskakis, and Carretero-Gonz\'alez provides
an excellent summary of this field~\cite{kevrekidisPG2008}.
The nonlinear mean field description given by the nonlinear
Schrodinger equation (NLSE) has been very accurate in the majority of experiments on BECs.  Vector
and non-local generalizations of the NLSE have also proven useful~\cite{kevrekidisPG2008}.  
In optical lattices, the mean field
description remains accurate provided the lasers creating the standing wave which is the optical lattice
are not too intense, and the dimensionality of the system is greater than one~\cite{carr2007h,carr2007g}.

In this article, we present a completely new class of nonlinear phenomena in BECs, based on the nonlinear
Dirac equation (NLDE).  Nonlinear Dirac equations have a long history in the literature, particularly in the
context of particle and nuclear theory~\cite{fushchich89,toyama94,liangzhong00,ngWK2007}, but also in applied mathematics and nonlinear dynamics~\cite{cazenave86,estebanMJ1995,fushchych97,esteban02,maccari05}. As nonlinearity is a ubiquitous aspect of Nature, it is natural to ask how nonlinearity might appear in a relativistic setting.  However, this line of questioning has been strongly constrained by modeling, rather than
first principles.  That is, there is no standard first principle of quantum electrodynamics (QED) which
is nonlinear.  So, the approach has been to require symmetry constraints in nonlinear models.  One of these
constraints is the principle of relativity, i.e., Poincar\'e covariance.  Poincar\'e symmetry includes
rotations, translations, and Lorentz boosts.

In contrast, our NLDE is not a model: it is derived from first principles for a weakly interacting bosonic gas in the presence
of a honeycomb optical lattice.  We show that Poincar\'e symmetry is naturally broken by the nonlinearity
inherent in this system.  Given that this form of nonlinearity, which depends only on the local
condensate density, is one
of the most common throughout nature, it is important to recognize that the principle of relativity may
be broken by small nonlinearities even at a fundamental level, for example of QED~\cite{ferrandoA2007,katsnelsonMI2007}.
Thus, we suggest a new direction of investigation
in  particle physics of possible nonlinearities as well as providing a natural context in artificial solid
state systems for the NLDE.

This article is outlined as follows.  In Sec.~\ref{sec:2} we provide a rigorous derivation of the NLDE from first principles.  In Sec.~\ref{sec:3} we discuss both discrete and continuous symmetries common to relativistic systems,
providing a clear physical interpretation in the present context.  Finally, in Sec.~\ref{sec:4} we conclude.

\section{ The nonlinear Dirac equation }
\label{sec:2}

\subsection{Two-Component Spinor Form of the NLDE}
\label{ssec:2component}

The second quantized Hamiltonian for a weakly interacting bosonic gas in two spatial dimensions is
 \begin{eqnarray}
 \hat{H} &=& \int d^2r\, \hat{\psi}^\dagger H_0 \hat{\psi} + \frac{g}{2} \int d^2r \,\hat{\psi}^\dagger\hat{\psi}^\dagger  \hat{\psi} \hat{\psi}\,,\label{eqn:H}\\
 H_0 &\equiv& -\frac{\hbar^2}{2m}\nabla^2 + V(\vec{r})\,.
 \end{eqnarray}
The bosonic field operators $\hat{\psi}=\hat{\psi}(\vec{r},t)$ obey bosonic commutation relations
in the Heisenberg picture.
In Eq.~(\ref{eqn:H}), $g\equiv 4 \pi \hbar^2 a_s / m$ is the coupling strength for binary contact interactions
with $a_s$ the s-wave scattering length and $m$ the atomic mass.  The external potential $V(\vec{r})$ is
a honeycomb lattice formed by standing waves of three sets of counter-propagating laser beams~\cite{zhuSL2007}.
The atoms experience this potential via the AC Stark effect.  We assume that the third spatial dimension
is frozen out by a tightly confining potential which is locally harmonic, as in Ref.~\cite{hadzibabic2006}.

\begin{figure}[t]
     \centering
     \subfigure[Hexagonal lattice structure]{
          \label{fig:1a}
          \includegraphics[width=7.5cm]{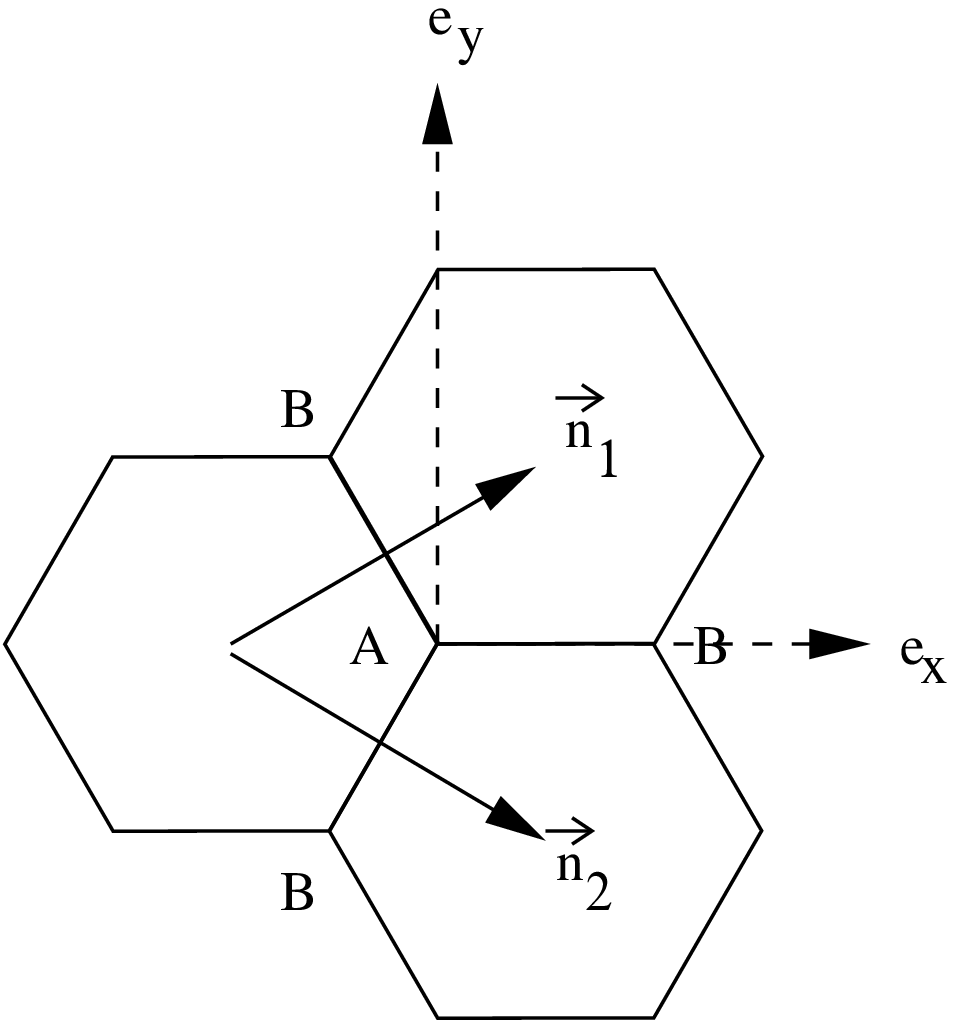}}\\
     \vspace{0.1in}
     \subfigure[Reciprocal lattice]{
           \label{fig:1b}
           \includegraphics[width=4.0cm]{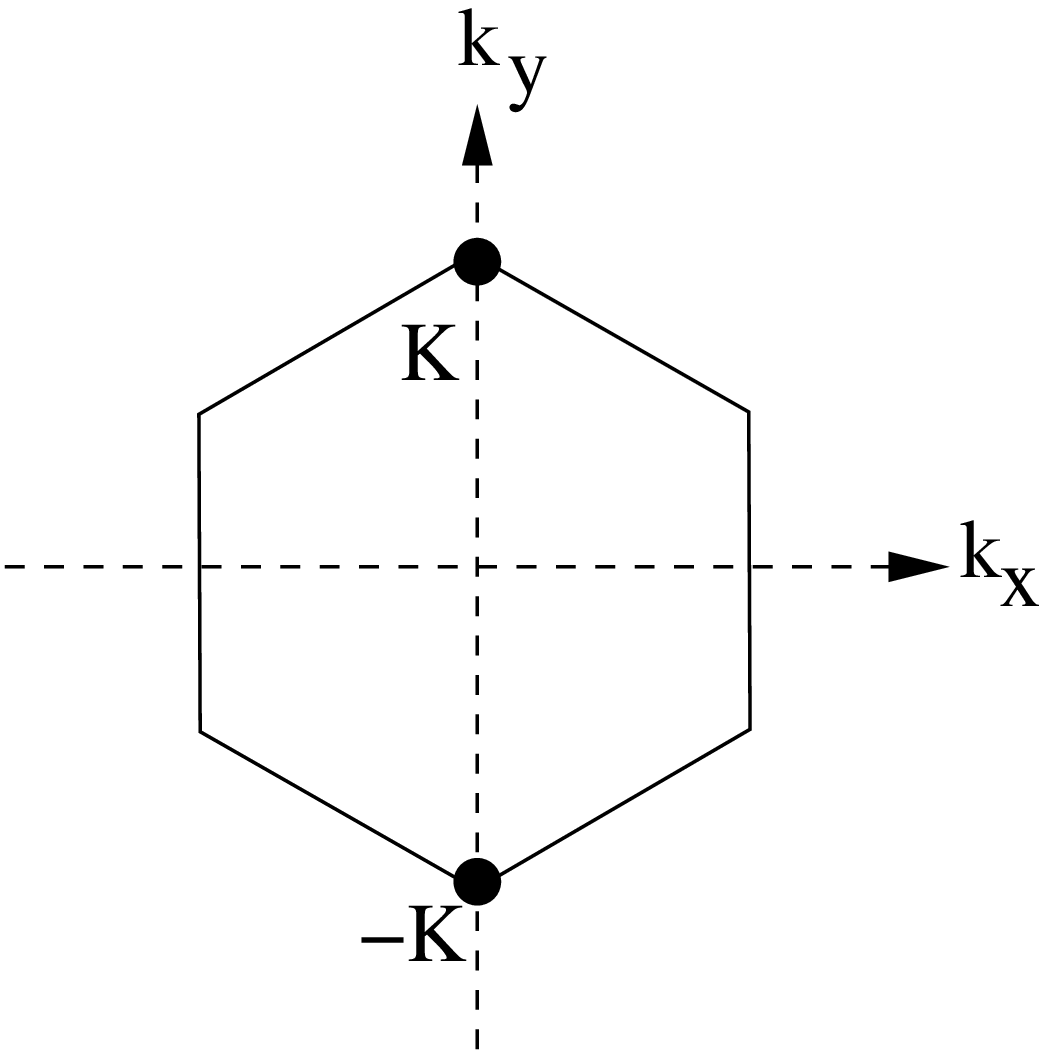}}
           \hspace{0.1in}
                \subfigure[Nearest neighbor displacement vectors]{
           \label{fig:1c}
           \includegraphics[width=4.0cm]{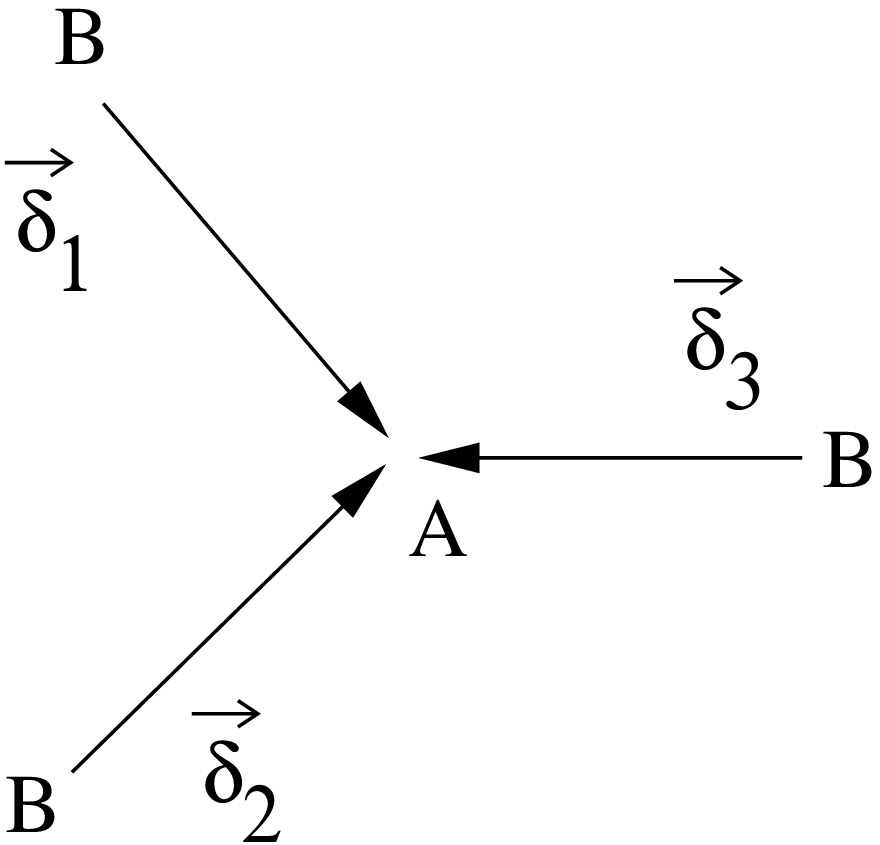}}
     \caption{Characterization of a honeycomb lattice.}
     \label{fig:1}
\end{figure}

The honeycomb lattice has two sites in the lattice unit cell.  We refer to the resulting two degenerate
sublattices as A and B.  Expanding in terms of Bloch states
in the lowest band 
belonging to A or B sites of the honeycomb lattice, as shown in Fig.~\ref{fig:1}, we can break up
the bosonic field operator into a sum over the two sublattices:
 \begin{eqnarray}
 \hat{\psi}&=& \hat{\psi}_A + \hat{\psi}_B  \,,\label{eqn:breakUpPsi}\\
 \hat{\psi}_A &\equiv& \sum_A \hat{a}\,  e^{i\vec{k}\cdot (\vec{r}-\vec{r}_A) } u( \vec{r}-\vec{r}_A)  \,,\label{eqn:psiA}\\
 \hat{\psi}_B &\equiv& \sum_B \hat{b}\,  e^{i\vec{k}\cdot (\vec{r}-\vec{r}_B) } u( \vec{r}-\vec{r}_B)\,,\label{eqn:psiB}
 \end{eqnarray}
where $\hat{a}$  and $ \hat{b}$  are the time-dependent destruction operators at A and B sites and $\vec{r}_A$ and $\vec{r}_B$ are the positions of A and B sites, respectively.
The spatial dependence is then encapsulated outside the operator in the exponential and the functions $u$.
The summation indices indicate sums over A or B sites.

Inserting Eq.~(\ref{eqn:breakUpPsi}) into Eq.~(\ref{eqn:H}), the Hamiltonian can be rewritten
 \begin{eqnarray}
 \hat{H} &=& \int d^2r    \left[(\hat{\psi}^{\dagger}_A + \hat{\psi}^{\dagger}_B) H_0 (\hat{\psi}_A + \hat{\psi}_B) \right.\nonumber  \\
 &&\left.+ \frac{g}{2}(\hat{\psi}^{\dagger}_A + \hat{\psi}^{\dagger}_B)(\hat{\psi}^{\dagger}_A + \hat{\psi}^{\dagger}_B) (\hat{\psi}_A + \hat{\psi}_B) (\hat{\psi}_A + \hat{\psi}_B)\right]\,.
 \label{eqn:H2}\end{eqnarray}
In the integral over $H_0$, imposing the restriction of  nearest-neighbor interactions in the tight-binding,
lowest band approximation eliminates all A-A and B-B transitions except for on-site kinetic and potential terms;
the latter can be neglected as an overall self-energy.  Then only integrals involving neighboring A-B sites remain in the sum.  Similarly, in the interaction term only
on-site terms are non-negligible, i.e., overlap of functions $u$ belonging to the same site.  Thus in the tight-binding, lowest band approximation, Eqs.~(\ref{eqn:psiA})-(\ref{eqn:psiB}) are substituted into Eq.~(\ref{eqn:H2}) to yield:
 \begin{eqnarray}
 \hat{H}&=& \int d^2r  \sum_{<A,B>}
 \left[ \hat{a}^{\dagger} \, e^{-i\vec{k}\cdot \vec{\chi}_A}
 u(\vec{\chi}_A) H_0 \hat{a}  \, e^{i\vec{k}\cdot \vec{\chi}_A } u(\vec{\chi}_A)\right.\nonumber \\
 &&+\hat{a}^{\dagger}\,  e^{-i\vec{k}\cdot \vec{\chi}_A } u(\vec{\chi}_A) H_0 \hat{b}\,  e^{i\vec{k}\cdot \vec{\chi}_B } u( \vec{\chi}_B) \nonumber\\
 &&+\hat{b}^{\dagger}\,  e^{-i\vec{k}\cdot \vec{\chi}_B } u( \vec{\chi}_B) H_0 \hat{a}\,  e^{i\vec{k}\cdot \vec{\chi}_A } u(\vec{\chi}_A) \nonumber \\
 &&\left.+\hat{b}^{\dagger}\,  e^{-i\vec{k}\cdot \vec{\chi}_B } u( \vec{\chi}_B) H_0 \hat{b}\,  e^{i\vec{k}\cdot \vec{\chi}_B } u( \vec{\chi}_B)\right] \nonumber \\
 &&+ \frac{g}{2} \int d^2r  \sum_A    \hat{a}^{\dagger} \hat{a}^{\dagger}  \hat{a}\hat{a}  \left[u( \vec{\chi}_A)\right]^4\nonumber\\
 &&+ \frac{g}{2}\int d^2r  \sum_B    \hat{b}^{\dagger} \hat{b}^{\dagger}  \hat{b}\hat{b} [u( \vec{\chi}_B)]^4\,,
\label{eqn:H3}\\
 \vec{\chi}_A&\equiv&\vec{r}-\vec{r}_A\,,\:\:\:\vec{\chi}_B\equiv\vec{r}-\vec{r}_B\,,\:\:\:
 \vec{\chi}_{AB}=\vec{r}_A - \vec{r}_B\,.
 \end{eqnarray}
Here the bracketed A and B summation index signifies a sum over nearest-neighbor A and B sites.  Isolating the integrals by pulling out all sums and terms not dependent on $\vec{r}$, Eq.~(\ref{eqn:H3}) becomes
 \begin{eqnarray}
 \hat{H} &=&  \sum_{<A,B>} \left[\hat{a}^{\dagger} \hat{b}\,e^{i\vec{k}\cdot \vec{\chi}_{AB} } \int d^2r e^{-i\vec{k}\cdot \vec{r} } u( \vec{\chi}_A)H_0  e^{i\vec{k}\cdot \vec{r} } u( \vec{\chi}_B)  \right.\nonumber \\
 &&+ \left.\hat{b}^{\dagger} \hat{a}\,e^{-i\vec{k}\cdot \vec{\chi}_{AB}} \int d^2r e^{-i\vec{k}\cdot \vec{r} } u( \vec{\chi}_B)H_0 e^{i\vec{k}\cdot \vec{r} } u( \vec{\chi}_A) \right]\nonumber \\
 &&+ \sum_{A} \hat{a}^{\dagger} \hat{a} \int d^2r e^{-i\vec{k}\cdot \vec{r} } u( \vec{\chi}_A)H_0  e^{i\vec{k}\cdot \vec{r} } u( \vec{\chi}_A)       \nonumber \\
 &&+ \sum_{B} \hat{b}^{\dagger} \hat{b} \int d^2r e^{-i\vec{k}\cdot \vec{r} } u( \vec{\chi}_B)H_0  e^{i\vec{k}\cdot \vec{r} } u( \vec{\chi}_B)       \nonumber \\
 &&+ \frac{g}{2}  \sum_A    \hat{a}^{\dagger} \hat{a}^{\dagger}  \hat{a}\hat{a} \int d^2r [u( \vec{\chi}_A)]^4 \nonumber\\ &&+ \frac{g}{2}   \sum_B  \hat{b}^{\dagger} \hat{b}^{\dagger}  \hat{b}\hat{b}  \int d^2r [u( \vec{\chi}_B)]^4 \,,
 \label{eqn:H4}\end{eqnarray}
Finally, we redefine the spatial integrals in Eq.~(\ref{eqn:H4}) as hopping energy  $t_h$ and interaction energy $U$, respectively, as is standard for the Hubbard Hamiltonian~\cite{reyAM2004,blakiePB2004} (note that we reserve
$t$ for time).  The terms in Eq.~(\ref{eqn:H4}) proportional
to $\hat{a}^{\dagger} \hat{a}$ and $\hat{b}^{\dagger} \hat{b}$ just count the total number of atoms in the system,
and can been neglected as an overall constant.  This leads to the Hamiltonian
 \begin{eqnarray}
 \hat{H}&=&   -t_h\sum_{<A,B>} \left[\hat{a}^{\dagger} \hat{b}\,e^{i\vec{k}\cdot (\vec{r}_A-\vec{r}_B) } +  \hat{b}^{\dagger} \hat{a}\,e^{-i\vec{k}\cdot (\vec{r}_A-\vec{r}_B) }\right] \nonumber \\
 &&+ \frac{U}{2}  \sum_A    \hat{a}^{\dagger} \hat{a}^{\dagger}  \hat{a}\hat{a} + \frac{U}{2}  \sum_B    \hat{b}^{\dagger} \hat{b}^{\dagger}  \hat{b}\hat{b}\,.  \label{eqn:Hubbard}
 \end{eqnarray}
Equation~(\ref{eqn:Hubbard}) is the Hubbard Hamiltonian divided into
two degenerate sublattices A and B, appropriate to the honeycomb optical lattice.

In order to work towards the nonlinear Dirac equation, we calculate the time evolution of $\hat{a}$ and $\hat{b}$ according to the standard Heisenberg picture prescription.  This is similar to the approach taken by
Pitaevskii in his landmark paper which first obtained the NLSE, or Gross-Pitaevskii equation~\cite{pitaevskii1961}.
The Heisenberg equation of motion is
 \begin{eqnarray}i\hbar \,\partial_t\hat{a}_k= [\hat{a}_k, \hat{H}]\,.\label{eqn:heisenberg}\end{eqnarray}
The operator $\hat{a}_k$, which destroys a boson at site $k$ on sublattice A,
satisfies the bosonic commutation relation \begin{equation}[\hat{a}_k,\hat{a}^{\dagger}_{k'}] = \delta_{kk'}\label{eqn:commutator}\,.\end{equation}
Then the commutator with the on-site interaction terms reduces to
 \begin{equation}[\hat{a}_k,\hat{a}^{\dagger}_k\hat{a}^{\dagger}_k \hat{a}_k\hat{a}_k  ] = \hat{a}_k \hat{a}^{\dagger}_k\hat{a}^{\dagger}_k \hat{a}_k\hat{a}_k -  \hat{a}^{\dagger}_k \hat{a}^{\dagger}_k\hat{a}_k \hat{a}_k\hat{a}_k. \label{eqn:commute2}
 \end{equation}
Taking the first product on the right and commuting the furthermost left $\hat{a}$ through according to Eq.~(\ref{eqn:commutator}), one finds:
 \begin{equation} \hat{a}_k \hat{a}^{\dagger}_k\hat{a}^{\dagger}_k \hat{a}_k\hat{a}_k = 2\hat{a}^{\dagger}_k \hat{a}_k\hat{a}_k + \hat{a}^{\dagger}_k \hat{a}^{\dagger}_k\hat{a}_k \hat{a}_k\hat{a}_k.
 \nonumber \label{eqn:commute3}\end{equation}
 Substituting Eq.~(\ref{eqn:commute3}) into Eq.~(\ref{eqn:commute2}), one obtains
 \begin{equation}
 [\hat{a}_k,\hat{a}^{\dagger}_k\hat{a}^{\dagger}_k \hat{a}_k\hat{a}_k ]  =   2\hat{a}^{\dagger}_k \hat{a}_k\hat{a}_k\,. \label{eqn:commute4}
 \end{equation}
Substituting Eq.~(\ref{eqn:commute4}) into Eq.~(\ref{eqn:heisenberg}) and the Hubbard Hamiltonian
Eq.~(\ref{eqn:Hubbard}), one finds
\begin{eqnarray} i\hbar \,\partial_t\hat{a}_k &=&  -  t_h \left[\hat{b}_k\,e^{i\vec{k}\cdot (\vec{r}_{A_k}-\vec{r}_{B_k}) } + \hat{b}_{k-n_1}\,e^{i\vec{k}\cdot (\vec{r}_{A_k}-\vec{r}_{B_{k-n_1}}) } \right.\nonumber \\
&&\left.+ \hat{b}_{k-n_2}\,e^{i\vec{k}\cdot (\vec{r}_{A_k}-\vec{r}_{B_{k-n_2}}) }\right] + U  \hat{a}^{\dagger}_k \hat{a}_k\hat{a}_k \label{eqn:heisenbergA}\end{eqnarray}
where the first three terms on the right hand side represent transitions from the three B-sites nearest the $k^{\mathrm{th}}$ site of the A sublattice and $\vec{n}_1$ and $\vec{n}_2$  are primitive cell translation vectors for the reciprocal lattice, as shown in Fig.~\ref{fig:1}.

In a similar fashion as Eqs.~(\ref{eqn:heisenberg})-(\ref{eqn:heisenbergA}), we arrive at an expression
of the same form for the B sublattice:
\begin{eqnarray}i\hbar\, \partial_t\hat{b}_k &=&  -   t_h \left[\hat{a}_k\,e^{-i\vec{k}\cdot (\vec{r}_{A_k}-\vec{r}_{B_k}) } +  \hat{a}_{k+n_1}\,e^{-i\vec{k}\cdot (\vec{r}_{A_{k+n_1}}-\vec{r}_{B_{k}}) } \right.\nonumber\\ &&\left.+ \hat{a}_{k+n_2}\,e^{-i\vec{k}\cdot (\vec{r}_{A_{k+n_2}}-\vec{r}_{B_{k}}) }\right] + U  \hat{b}^{\dagger}_k \hat{b}_k\hat{b}_k .   \label{eqn:heisenbergB}\end{eqnarray}
Continuing to follow Pitaevskii's method, we next take the expectation value of Eqs.~(\ref{eqn:heisenbergA}) and~(\ref{eqn:heisenbergB}) with respect to on-site coherent states.
A tensor product over sites of such coherent states is also assumed~\cite{carr2007h}.  A more formal, careful treatment of finite number states, rather than coherent states, has been worked out in the literature (see~\cite{castinY2001} and references therein).  Either way, we obtain coupled equations of motion for discrete, on-site, complex-valued amplitudes.
For simplicity of notation we take
 \begin{equation} a_k \equiv \langle\hat{a}_k\rangle\,,\:\:\:b_k \equiv \langle\hat{b}_k\rangle\,.
 \end{equation}
Inserting the nearest-neighbor vectors $\vec{\delta}_1$, $\vec{\delta}_2$, and $\vec{\delta}_3$ in the exponentials in Eqs.~(\ref{eqn:heisenbergA}) and~(\ref{eqn:heisenbergB}),
as shown in Fig.~\ref{fig:1}, we obtain
 \begin{eqnarray}i\hbar \,\dot{ {a}}_k &=&  -  t_h ({b}_k\,e^{i\vec{k}\cdot \vec{\delta}_3 } + {b}_{k-n_1}\,e^{i\vec{k}\cdot \vec{\delta}_1}  + {b}_{k-n_2}\,e^{i\vec{k} \cdot \vec{\delta}_2}   )\nonumber\\
 &&+ U a^*_ka_ka_k \,, \label{eqn:heisA2}\\
 i\hbar \, \dot{ {b}}_k &=& -   t_h ({a}_k\,e^{-i\vec{k}\cdot \vec{\delta}_3} + {a}_{k+n_1}\,e^{-i\vec{k}\cdot \vec{\delta}_1 }  + {a}_{k+n_2}\,e^{-i\vec{k}\cdot \vec{\delta}_2}    ) \nonumber \\
 &&+ U b^*_kb_kb_k \,, \label{eqn:heisB2}\end{eqnarray}
where $n_1$ and $n_2$ in the indices label the lattice sites in the two directions of the primitive-cell translation vectors $\vec{n}_1$ and $\vec{n}_2$.

The NLDE is derived around the linear band crossings between the A and B sublattices at the Brillouin zone
corners~\cite{wallacePR1947}, called
\emph{Dirac cones} in the graphene literature~\cite{geim2007}.  To this end, we insert particular values for the nearest-neighbor displacement vectors $\vec{\delta}$ and evaluate $\vec{k}$ at the Brillouin zone corner, defined by $\vec{k} = \vec{K} = (0,4\pi/3)$ , $\vec{\delta}_1=(\frac{1}{2\sqrt{3} }, -\frac{1}{2 })$, $ \vec{\delta}_2=(\frac{1}{2\sqrt{3} }, \frac{1}{2 })$, $\vec{\delta}_3=(-\frac{1}{\sqrt{3}},0)$, all in the $x-y$ plane.  Then Eq.~(\ref{eqn:heisA2}) becomes
 \begin{eqnarray}
 i\hbar \,\dot{ {a}}_k &=&  -  t_h ({b}_k\,e^0 +  {b}_{k-n_1}\,e^{-i2\pi/3}  + {b}_{k-n_2}\,e^{i2\pi/3}   )\nonumber\\
 &&+ U a^*_ka_ka_k\,.\label{eqn:heisA3}
 \end{eqnarray}
Reducing the exponentials,
 \begin{eqnarray}
 i\hbar  \,\dot{ {a}}_k &=&  -   t_h [{b}_k + {b}_{k-n_1}(-1/2 - i \sqrt{3}/2)\nonumber\\
 &&+ {b}_{k-n_2}(-1/2 +i \sqrt{3}/2)]+ U a^*_ka_ka_k\,.\label{eqn:heisA3_5}\end{eqnarray}

In anticipation of taking the long wavelength, continuum limit, as is necessary to obtain the NLDE, we group terms appropriately in Eq.~(\ref{eqn:heisA3_5}) in order to construct discrete versions of derivatives:  \begin{eqnarray}i\hbar  \,\dot{ {a}}_k  &=&   -  t_h [{b}_k + (b_k - {b}_{k-n_1})(1/2 +i \sqrt{3}/2) \nonumber\\
 &&- b_k (1/2 +i \sqrt{3}/2) +(b_k - {b}_{k-n_2})(1/2 - i\sqrt{3}/2) \nonumber\\
 &&- b_k (1/2 - i\sqrt{3}/2)]  +U a^*_ka_ka_k\,,
 \end{eqnarray}
which reduces to
 \begin{eqnarray} i\hbar  \dot{ {a}}_k &=&  -  t_h [ (b_k - {b}_{k-n_1})(1/2 + i\sqrt{3}/2) \nonumber\\
 &&+(b_k - {b}_{k-n_2})(1/2 - i\sqrt{3}/2)]+ U a^*_ka_ka_k\,.
 \end{eqnarray}
Taking the continuum limit and replacing the discrete quantities $a_k=a_k(t)$ and $b_k=b_k(t)$ by the continuous functions $\psi_A=\psi_A(\vec{r},t)$ and $\psi_B=\psi_B(\vec{r},t)$, we arrive at
 \begin{eqnarray}
 i\hbar \dot{ {\psi}_A} &=&  -   t_h \left[\partial_{n_1} \psi_B (1/2 + i\sqrt{3}/2) 
 +\partial_{n_2}(1/2 - i\sqrt{3}/2)\right] \nonumber \\
 &&+ U \psi^*_A \psi_A \psi_A \,.\label{eqn:heisA4}
 \end{eqnarray}
where the partial derivatives are in the directions of the unit-cell vectors $\vec{n}_1$ and  $\vec{n}_2$.  With a little trigonometry we find that the unit-cell vectors are
 \begin{eqnarray} \vec{n}_1 &=& \cos(\pi/6) \hat{e}_x - \sin(\pi/6) \hat{e}_y= \sqrt{3}/2 \hat{e}_x - 1/2 \hat{e}_y\,,\label{eqn:n1}\\
 \vec{n}_2 &=& \cos(\pi/6) \hat{e}_x + \sin(\pi/6) \hat{e}_y = \sqrt{3}/2 \hat{e}_x + 1/2 \hat{e}_y\,.\label{eqn:n2} \end{eqnarray}
Up to now the ``hat'' symbol (circumflex accent) has been reserved for operators alone.  However, in Eqs.~(\ref{eqn:n1})-(\ref{eqn:n2}) we use this symbol to indicate a unit vector in the $x$ and $y$ directions.  Derivatives with respect to the unit-cell vectors take the form
 \begin{eqnarray} \partial_{n_1} &=&    \vec{n_1}  \cdot   \vec{\nabla} =  (\sqrt{3}/2 ) \partial_x - (1/2) \partial_y  \,,\label{eqn:partial1}\\
 \partial_{n_2} &=&    \vec{n_2}  \cdot   \vec{\nabla} =  (\sqrt{3}/2) \partial_x + (1/2 ) \partial_y\,.  \label{eqn:partial2}
 \end{eqnarray}
Substituting Eqs.~(\ref{eqn:partial1})-(\ref{eqn:partial2}) into Eq.~(\ref{eqn:heisA4}),
 \begin{eqnarray}  i\hbar\, \dot{ {\psi}_A}&=& - t_h \left\{ [(\sqrt{3}/2) \partial_x- (1/2) \partial_y]     \psi_B[(1/2) +i (\sqrt{3}/2)]  \right.\nonumber\\
 &&\left.+ [(\sqrt{3}/2) \partial_x + (1/2) \partial_y]  \psi_B  [(1/2) - i(\sqrt{3}/2)]\right\}\nonumber\\
 &&+ U \psi^*_A \psi_A \psi_A\,. \label{eqn:heisA5}
 \end{eqnarray}
Further simplification of Eq.~(\ref{eqn:heisA5}) leads to
\begin{equation} i\hbar \dot{\psi}_A=-    \frac{t_h\sqrt{3}}{2} (  \partial_x \psi_B - i   \partial_y \psi_B )   + U \psi^*_A \psi_A \psi_A\,.\label{eqn:heisAFinal}\end{equation}
Similarly, for the continuum limit of $b_k = b_k(t) \rightarrow \psi_B = \psi_B(\vec{r},t)$,
 \begin{eqnarray} i\hbar \dot{\psi}_B =  -    \frac{t_h\sqrt{3}}{2} ( -  \partial_x \psi_A - i   \partial_y \psi_A )  + U \psi^*_B \psi_B \psi_B\,. \label{eqn:heisBFinal}
 \end{eqnarray}

Equations~(\ref{eqn:heisAFinal})-(\ref{eqn:heisBFinal}) are in fact massless Dirac equations with an added
nonlinear term.  To put this in a more familiar form, we use Pauli matrix notation.  To this end, one must rename the coordinate axes so that $x \rightarrow y$, and, in order to preserve the handedness of the coordinate system, $y\rightarrow -x$.  We also reinsert the lattice constant $a$; note that $a$ is unrelated
to the s-wave scattering length $a_s$ briefly mentioned in the definition of $g$ following Eq.~(\ref{eqn:H}).
Thus $\partial_x \rightarrow \partial_y$ and $ \partial_y \rightarrow -\partial_x$.  Then Eqs.~(\ref{eqn:heisAFinal})-(\ref{eqn:heisBFinal}) become
 \begin{eqnarray} i\hbar \dot{\psi}_A &=&-   \frac{t_h a\sqrt{3}}{2} ( \partial_y \psi_B + i  \partial_x \psi_B )   + U \psi^*_A \psi_A \psi_A \,,\\
 i\hbar \dot{\psi}_B &=& -  \frac{t_h a \sqrt{3}}{2} ( -\partial_y \psi_A + i \partial_x \psi_A)  + U \psi^*_B \psi_B \psi_B\,,
 \end{eqnarray}
or in matrix form,
 \begin{eqnarray}
 i\hbar \left[
  \begin{array}{ c c }
     \dot{\psi}_A \\
     \dot{\psi}_B
  \end{array} \right]
&=& \frac{  -it_ha \sqrt{3}}{2}  \left[
  \begin{array}{ c c }
     0 &     \partial_x -i \partial_y  \\
     \partial_x + i \partial_y   & 0
  \end{array} \right]  \left[
  \begin{array}{ c c }
     \psi_A \\
     \psi_B
  \end{array} \right] \nonumber \\
  &&+ U  \left[
  \begin{array}{ c c }
     \psi_A^*\psi_A\psi_A \\
     \psi_B^*\psi_B\psi_B
  \end{array} \right] \,.\label{eqn:nlde1}\end{eqnarray}
We can write Eq.~(\ref{eqn:nlde1}) more compactly in terms of Pauli matrices $(\sigma_x,\sigma_y)=\vec{\sigma}$,
  \begin{eqnarray}
  i\hbar \left[
  \begin{array}{ c c }
     \dot{\psi}_A \\
     \dot{\psi}_B
  \end{array} \right]
= \frac{  -it_h a \sqrt{3}}{2} \vec{ \sigma} \cdot  \vec{\nabla }    \left[
  \begin{array}{ c c }
     \psi_A \\
     \psi_B
  \end{array} \right] + U  \left[
  \begin{array}{ c c }
     \psi_A^*\psi_A\psi_A \\
     \psi_B^*\psi_B\psi_B
  \end{array} \right]\!.\label{eqn:nlde2}
  \end{eqnarray}
Equation~(\ref{eqn:nlde2}) is the NLDE in (2+1) dimensions.

However, we can make one further step by expressing Eq.~(\ref{eqn:nlde2})
in a more covariant-looking form as follows:
 \begin{equation}
 (i \sigma_0 \partial_t   +i c_s\vec{\sigma}\cdot \vec{\nabla}) \left[
 \begin{array}{ c c }
 \psi_A \\
 \psi_B
 \end{array} \right]
 - U  \left[
 \begin{array}{ c c }
 \psi_A^*\psi_A\psi_A \\
 \psi_B^*\psi_B\psi_B
 \end{array} \right] = 0 \,,\label{eqn:nlde3}
 \end{equation}
where $\vec{\sigma}$ and $\vec{\nabla}$ are restricted to the $x-y$ plane.
In Eq.~(\ref{eqn:nlde3}) \begin{equation}c_s \equiv   \frac{t_h a\sqrt{3}}{2\hbar}\label{eqn:soundspeed}\end{equation} is an effective speed of sound of the condensate
in the lattice\cite{taylorE2003,wuBiao2006} (in graphene it would be replaced with the Fermi velocity~\cite{geim2007}).  This velocity is an effective speed of light for excitations of the NLDE in our QED$_{2+1}$ theory.  Experimental values of $c_s$ in BECs are on the order of cm/s, ten orders of magnitude slower than the speed of light in a vacuum.  Note also that in Eq.~(\ref{eqn:nlde3}) $U$ has now absorbed a factor of $1/\hbar$.

Finally, a few additional definitions lead to a nicely compact form for the NLDE.  Let
\begin{eqnarray}A\equiv\!  \left[
  \begin{array}{ c c }
     1 &     0  \\
    0   & 0
  \end{array} \right]\!,   B \equiv\!  \left[
  \begin{array}{ c c }
     0 &     0  \\
    0   & 1
  \end{array} \right]\!,
  \psi\equiv\!\left[
  \begin{array}{ c c }
     \psi_A \\
     \psi_B
  \end{array} \right]\!,
  \bar{\psi}\equiv\!\left[
  \begin{array}{ c c }
     \psi_A^*
     ,\psi_B^*
  \end{array} \right]\!.  \label{eqn:AandB}\end{eqnarray}
With the choice of metric which raises and lowers space-time indices restricted to (2+1) dimensions,
  \begin{eqnarray}  g^{\mu \nu} =   \left[
  \begin{array}{ c c c}
     1 &  0 & 0 \\
    0  & -1 & 0 \\
    0 & 0 & -1
  \end{array} \right]\,,\end{eqnarray}
Eq.~(\ref{eqn:nlde3})  becomes
\begin{eqnarray}
( i \sigma^\mu \partial_\mu - U \bar{\psi } A \psi A   -   U \bar{\psi } B \psi B  ) \psi = 0\,, \label{eqn:nldeFinal}
\end{eqnarray}
where the standard Einstein summation rule is in effect, $\mu \in \{ 0,1,2\}$ in keeping with (2+1) dimensions,
and the units are chosen such that $c_s=1$.

\subsection{Maximally Compact Form of the NLDE}
\label{ssec:compact}

In Sec.~\ref{ssec:2component} we developed $\psi$, a two-dimensional complex object which brings to mind one member of a pair of Weyl-spinors in the (1/2,1/2) chiral representation of the Dirac algebra used to describe massless neutrinos in the standard QED$_{3+1}$ theory.  Such a treatment is appropriate for any neutral Dirac fermion viewed in the extreme relativistic frame.  In order to make the connection clear we must find the second member of the pair
of Weyl-spinors and verify that the mapping is true. A thorough treatment of the mapping of QED$_{3+1}$ into the QED$_{2+1}$ theory of graphene is contained in \cite{gusyninVP2007} and references therein.  We will continue to
restrict ourselves to (2+1) dimensions in the following.

To this end, we seek to put the NLDE into a form consistent with the standard compact four-component spinor notation for the linear Dirac equation
 \begin{equation}i \gamma^{\mu} \partial_{\mu}\Psi = 0\,,\label{eqn:dirac}
 \end{equation}
where the $\gamma^{\mu}$ are the $4\times 4$ Dirac matrices~\cite{bjorken64}.  We point out that this notation is not only standard but more appropriate if the quasi-particles, i.e., the long-wavelength excitations, develop a non-zero effective mass, as can be caused by lattice distortion~\cite{zhuSL2007}.

We obtained the NLDE by evaluating the exponentials at the Brillouin Zone corner $\vec{K}_+= (0,4\pi/3)$.  There is another inequivalent corner,  as shown in Fig.~\ref{fig:1}, $\vec{K}_-\equiv (0,-4\pi/3)$, near which perturbations in momentum, i.e., long-wavelength quasiparticle excitations, are governed by a similar first-order wave equation.  By considering
equations of motions derived around \emph{both} Brillouin corners, we can obtain the four-vector notation.  The coupled equations evaluated at $\vec{K}_-$ are
\begin{eqnarray} i\hbar \dot{\psi}_A &=& \frac{  -t_h a\sqrt{3}}{2} ( \partial_x  + i \partial_y )\psi_B   + U \psi^*_A \psi_A \psi_A\,,\\
i\hbar \dot{\psi}_B &=& \frac{  -t_h a \sqrt{3}}{2} ( -\partial_x  + i \partial_y )\psi_A  + U \psi^*_B \psi_B \psi_B\,. \end{eqnarray}
Following the same steps as before we obtain
\begin{eqnarray}
  i\hbar \left[
  \begin{array}{ c c }
     \dot{\psi}_B \\
     \dot{\psi}_A
  \end{array} \right]
=    \frac{it_ha \sqrt{3}}{2} \vec{ \sigma} \cdot  \vec{\nabla }    \left[
  \begin{array}{ c c }
     \psi_B \\
     \psi_A
  \end{array} \right] + U  \left[
  \begin{array}{ c c }
     \psi_B^*\psi_B\psi_B \\
     \psi_A^*\psi_A\psi_A
  \end{array} \right]\!.\label{eqn:othercorner}
\end{eqnarray}
We combine Eqs.~(\ref{eqn:othercorner}) and~(\ref{eqn:nlde2}) into one equation involving a single 4-component object and attach $\pm$ subscripts to the wave functions $\Psi$ to specify the corner of the Brillouin Zone.  The resulting
expression is
   \begin{eqnarray}
  i \partial_t  \left[\begin{array}{ c  }
      \Psi_{+} \\
      \Psi_{- }
  \end{array} \right]
  &+& ic_s  \left[
  \begin{array}{ c c c}
      \bf{\vec{\sigma} \cdot \vec{\nabla}} &  \bf{ 0} \\
       \bf{0}   &  -\vec{\sigma} \cdot \vec{\nabla}
  \end{array} \right]  \left[
  \begin{array}{ c  }
      \Psi_{+} \\
      \Psi_{- }
  \end{array} \right]  \nonumber\\
   &&- U  \left[
  \begin{array}{ c }
      N_+ \\
      N_-
  \end{array} \right]    = 0 \,.\label{eqn:nlde4vec}
  \end{eqnarray}
where the nonlinear terms are grouped into the 2-vectors $N_+$ and $N_-$ defined by
  \begin{eqnarray}
N_+= \left[
  \begin{array}{ c }
       ( \psi_A^*\psi_A\psi_A)_+\\
       ( \psi_B^*\psi_B\psi_B)_+
  \end{array} \right]  ,\:\:\:
  N_-= \left[
  \begin{array}{ c }
       ( \psi_B^*\psi_B\psi_B)_-\\
       ( \psi_A^*\psi_A\psi_A)_-
  \end{array} \right]\,  \label{eqn:N+N-}
\end{eqnarray}
and the four-spinors are given by
 \begin{eqnarray}
 \Psi &\equiv& \left[  \begin{array}{ c  }
      \Psi_{+} \\
      \Psi_{- }
  \end{array} \right]
  \equiv
 \left[    \begin{array}{ c  }
      \psi_{A+} \\
      \psi_{B+} \\
      \psi_{B-} \\
      \psi_{A-}
  \end{array} \right] \,, \\
 \Psi^\dagger &\equiv& \left[
  \begin{array}{ c c  }
     \Psi_{+}^* & \Psi_{-}^*\\
  \end{array} \right]
  \equiv \left[
  \begin{array}{ c c c c }
     \psi_{A+}^* & \psi_{B+}^* & \psi_{B-}^* & \psi_{A-}^*\\
  \end{array} \right]\,.
  \label{eqn:4spinorDefs}\end{eqnarray}
Here again the $\pm$ subscripts refer to the specific corner of the Brillouin zone.

We can reduce Eq.~(\ref{eqn:nlde4vec}) to a more compact form by introducing the following $4 \times 4$ matrices:
\begin{eqnarray} {\Sigma}_0&=& \left[
  \begin{array}{ c c c}
      \bf{1}  &  \bf{ 0} \\
       \bf{0}   &  \bf{1}
  \end{array} \right]       ,   { \Sigma}_1= \left[
  \begin{array}{ c c c}
      \sigma_x &  \bf{ 0} \\
       \bf{0}   &  -\sigma_x
  \end{array} \right]             ,   { \Sigma}_2= \left[
  \begin{array}{ c c c}
      \sigma_y &  \bf{ 0} \\
       \bf{0}   &  -\sigma_y
  \end{array} \right] ,\label{eqn:def1}\\
   \it{A_+} &=& \left[
  \begin{array}{ c c c}
      \bf{A}  &  \bf{ 0} \\
       \bf{0}   &  \bf{0}
  \end{array} \right]         ,      \it{A_-} = \left[
  \begin{array}{ c c c}
      \bf{0}  &  \bf{ 0} \\
       \bf{0}   &  \bf{A}
  \end{array} \right]            ,    \\
  \it{B_+} &=& \left[
  \begin{array}{ c c c}
      \bf{B}  &  \bf{ 0} \\
       \bf{0}   &  \bf{0}
  \end{array} \right]              ,      \it{B_-} = \left[
  \begin{array}{ c c c}
      \bf{0}  &  \bf{ 0} \\
       \bf{0}   &  \bf{B}
  \end{array} \right]\,.\label{eqn:def3}
  \end{eqnarray}
The boldface notation ${\bf A}$ and ${\bf B}$ denote the 2$\times$2 matrices defined in  Eq.~(\ref{eqn:AandB}). Also, the boldface entries ${\bf 1}$ and ${\bf 0}$ in the matrices in  Eq.~(\ref{eqn:N+N-}) refer to the 2$\times$2 unit matrix and zero matrix, respectively~\cite{ngWK2007}.
Then Eq.~(\ref{eqn:nlde4vec}) becomes
 \begin{eqnarray} (i {\it  \Sigma^{\mu}}\partial_{\mu}  &-& U \Psi^\dagger {\it A_+} \Psi {\it A_+}   -   U \Psi^\dagger {\it B_+ }\Psi {\it B_+}  \nonumber \\  &-& U \Psi^\dagger {\it A_-} \Psi {\it A_- }  -   U \Psi^\dagger {\it B_-} \Psi {\it B_- })  \Psi = 0\,.   \label{eqn:nlde4vec2}\end{eqnarray}
We substitute into Eq.~(\ref{eqn:nlde4vec2}) the Dirac matrices in the Chiral representation:
 \begin{equation}\gamma^0 \equiv \left[
  \begin{array}{ c c c}
      \bf{0}  &  \bf{ 1} \\
       \bf{1}   &  \bf{0}
  \end{array} \right]        ,
  \gamma^1 \equiv \left[
  \begin{array}{ c c c}
      \bf{0}  &  -\sigma_x \\
       \sigma_x   &  \bf{0}
  \end{array} \right]     ,
  \gamma^2 = \left[
  \begin{array}{ c c c}
      \bf{0}  &  -\sigma_y \\
       \sigma_y   &  \bf{0}
  \end{array} \right]
  \label{eqn:diracMatrices}
  \end{equation}
and multiply on the left by $\gamma^0$.  Note that $\mu\in\{0,1,2\}$ in keeping with (2+1) dimensions. 
Finally, for the nonlinear term we introduce
  \begin{equation} N \equiv - U \Psi^\dagger \sum_{  Q \in {A_+,A_-,B_+,B_-}}{\it   Q} \Psi {\it   Q}. \nonumber \end{equation}
Implementing Dirac matrix notation as described, the NLDE of Eq.~(\ref{eqn:nlde4vec2}) becomes
  \begin{equation}
  (i   \gamma^{\mu}\partial_{\mu}  + \gamma^0  N)  \Psi = 0\,.\label{eqn:nlde4vec3}
 \end{equation}
Equation~(\ref{eqn:nlde4vec3}) is the most compact form of the massless NLDE in 4-component spinor notation.

\section{Symmetries and Constraints}
\label{sec:3}

The NLDE as expressed by Eq.~(\ref{eqn:nlde3}) or Eq.~(\ref{eqn:nldeFinal}) looks like the Dirac equation for a two-spinor, as stated in the graphene problem~\cite{geim2007}, with the addition of two nonlinear on-site interaction terms, one for each A and B sublattice; similarly, the more compact form developed in Sec.~\ref{ssec:compact} also appears to be a Dirac equation for a four-component spinor, with an additional term.  However, one should not be too hasty in assigning characteristics based on appearances.  We therefore make a careful and thorough exploration of the symmetries and other important mathematical properties of the NLDE.

In what follows we follow a similar route as in Ref.~\cite{ngWK2007}.  First we check the linear version with the delta interactions turned off to ensure that Eq.~(\ref{eqn:nldeFinal}) is indeed the massless Dirac equation in (2+1) dimensions with all the necessary symmetries.  For each symmetry we then check the nonlinear interaction terms to determine whether they preserve or break the symmetry.

\subsection{ Locality}
\label{ssec:3.1}

We do not necessarily require the evolution of the wavefunction as described by an NLDE to be governed by a local theory.  A local theory is one in which the terms in the linear equations of motion involve only factors of the wavefunction and its derivatives evaluated at the same space-time coordinate.  Nonlocality arises in low energy limits of some quantum field theories.  However, the nonlinearity in our NLDE is manifestly local.  Thus our NLDE is closer to the standard Dirac equation on the classical level (no quantum effects), modified by the on-site interaction term.  In Sec.~\ref{sec:4} we discuss the possibility of non-local nonlinearities, including for graphene.

\subsection{Poincar\'e Symmetry}
\label{ssec:3.2}

A Poincar\'e transformation takes the spatiotemporal point defined by the the 4-vector $r_{\nu}$ into the point $r'_{\mu}$ according to
 \begin{equation}r'_{\mu} = \Lambda_{\mu}^{\nu} r_\nu + d_{\mu}
 \end{equation}
where $\Lambda_{\mu}^{\nu} $ is the coordinate matrix representation of the Lorentz group and $d_{\mu}$ is a space-time translation.
The wave function $\Psi$ transforms as  \begin{eqnarray}\Psi'({\it r' }) = M({\it \Lambda}) \Psi ({\it r})\nonumber \end{eqnarray}
where the matrices $M({\it \Lambda}) $ form a representation of the subgroup of the Lorentz group consisting of spatial rotations and boosts.  Boosts can be thought of as rotations in imaginary time by imaginary angles mixing space and time coordinates.  We restrict these transformations to (2+1) dimensions.

The proof of Lorentz covariance of the standard massless Dirac equation, Eq.~(\ref{eqn:dirac}),  is arrived at with the aid of the transformations for the wave function and partial derivatives:
 \begin{eqnarray}\Psi({\it r}) &=& M^{-1}({\it \Lambda} )\Psi'({\bf r'})\,,\\
 \partial_{x_i} &=& \Lambda^{j}_{i} \partial_{x_j'}\,.
 \end{eqnarray}
This yields the conditions for the Dirac matrices:
 \begin{eqnarray} \gamma_j = \Lambda_{ji} M \gamma_iM^{-1}\,.
 \end{eqnarray}
The standard form of the Dirac matrices obtained this way can be found in the literature~\cite{bjorken64} and are identical to the results of our theory.

Thus imposing Lorentz covariance on the NLDE as expressed in Eq.~(\ref{eqn:nlde4vec3})
requires  $\Psi$ to transform under the irreducible representation of a subgroup of SL(2,C), the $2 \times 2$ complex matrices of unit determinant.
The four-dimensional representation of SL(2,C),  $D^{(1/2,1/2)}$, is formed by taking the direct product of the two-dimensional representations $D^{(1/2,0)}$and $ D^{(0,1/2)}$:
 \begin{eqnarray} D^{(1/2,1/2)} =D^{(1/2,0)} \otimes  D^{(0,1/2)}\,.
 \end{eqnarray}
This subgroup of SL(2,C) is isomorphic to a subgroup of the Lorentz group, the one obtained by restricting Lorentz transformations to the plane of the honeycomb lattice.  So, the upper two components of $\Psi$ transform as a spinor under $D^{(1/2,0)}$, the lower two as a spinor under $D^{(0,1/2)} $, and $\Psi $ itself as a four-component spinor or ${\it bispinor}$~\cite{hladikJ1999}.

The next task is to examine the behavior of $\Psi$, as defined in Eq.~(\ref{eqn:4spinorDefs}) and governed by Eq.~(\ref{eqn:nlde4vec3}), under rotations in the $x-y$ plane.  In order to obtain Poincar\'e covariance we must show that it is the same as that of a four-component spinor in the standard Dirac$_{3+1}$ theory restricted to the 2D plane.
The honeycomb lattice is invariant under rotations by $\pm 2\pi /3$ but the four components of $ \Psi$ are also defined by the particular corner of the Brillouin Zone.  Since we're considering a discrete lattice it is natural to discuss discrete rotations which realign lattice points and in the continuum limit map the continuous rotations of QED$_{2+1}$ onto our theory.
Rotations by $\pm \pi /3$ exchange A and B sites, and take the theory to that of the opposite $\vec{K}$ point:  $\vec{K}_+$ does not go to $\vec{K}_-$ but the result after calculating the relative phase exponentials gives back the same theory.  To see this we chose a different primitive unit cell, the one obtained by a rotation of $2\pi/3$ about the $\vec{n}_1 , \vec{n}_2$ origin in Fig~\ref{fig:1}.  This is because the direction of $\vec{K}_+^\prime$ (defined as $\vec{K}_+$ rotated by $\pi/3 $) differs from that of $\vec{K}_-$ by $2\pi/3$.

Thus under this discrete rotation
 \begin{eqnarray}
 b_k &\rightarrow&  b_{k-2n_1}\,, \label{eqn:firstPrimeSub}\\
 b_{k-n_1} &\rightarrow&  b_{k-n_1-n_2}\,,\\
 b_{k-n_2} &\rightarrow&  b_{k-n_1}\,,\\
 a_k &\rightarrow&  a_{k-n_1}\,,\\
 a_{k+n_1} &\rightarrow&  a_{k-2n_1+n_2}\,,\\
 a_{k+n_2} &\rightarrow&  a_{k-2n_1}\,.
 \end{eqnarray}
Also we observe that
\begin{eqnarray} \vec{K}_+^\prime\cdot \vec{\delta}_3  = \vec{K}_-\cdot \vec{\delta}_1    \\
\vec{K}_+^\prime\cdot \vec{\delta}_1  = \vec{K}_-\cdot \vec{\delta}_2    \\
 \vec{K}_+^\prime\cdot \vec{\delta}_2  = \vec{K}_-\cdot \vec{\delta}_3 \label{eqn:lastPrimeSub}
 \end{eqnarray}
Substituting Eqs.~(\ref{eqn:firstPrimeSub})-(\ref{eqn:lastPrimeSub}) into Eqs.~(\ref{eqn:heisA2})-(\ref{eqn:heisB2}) we obtain
 \begin{eqnarray}
 i\hbar \,\dot{a}_{k-n_1} &=&   -t_h ({b}_{k-2n_1}\,e^{i\vec{K}_-\cdot \vec{\delta}_1 } +{b}_{k-n_1-n_2}\,e^{i\vec{K}_- \cdot \vec{\delta}_2 } \nonumber \\
 &&+ {b}_{k-n_1}\,e^{i\vec{K}_-\cdot \vec{\delta}_3 }   ) + U a^*_{k-n_1}a_{k-n_1}a_{k-n_1}\,,
 \label{eqn:aRotate1}\\
 i\hbar \,\dot{b}_{k-2n_1} &=&   -t_h ({a}_{k-n_1}\,e^{-i\vec{K}_-\cdot \vec{\delta}_1} +{a}_{k-2n_1+n_2}\,
 e^{-i\vec{K}_-\cdot \vec{\delta}_2 }  \nonumber \\
 &&\!\!\!\!+ {a}_{k-2n_1}\,e^{-i\vec{K}_-\cdot \vec{\delta}_3  }) + U b^*_{k-2n_1}b_{k-2n_1}b_{k-2n_1}.
 \label{eqn:bRotate1}
 \end{eqnarray}
Now we redefine the index $k$: in Eq.~(\ref{eqn:aRotate1}), $k \rightarrow k + n_1$, while in Eq.~(\ref{eqn:bRotate1}), $k \rightarrow k + 2n_1$.  Then
 \begin{eqnarray}
 i\hbar \,\dot{a}_{k} &=&   -t_h ({b}_{k-n_1}\,e^{i\vec{K}_-\cdot \vec{\delta}_1 } +
 {b}_{k-n_2}\,e^{i\vec{K}_-\cdot \vec{\delta}_2 }
 \nonumber \\
 &&+ {b}_{k}\,e^{i\vec{K}_-\cdot \vec{\delta}_3 })+ U a^*_{k}a_{k}a_{k}  \label{eqn:aRotate2}\\
 i\hbar \,\dot{b}_{k} &=&   -t_h ({a}_{k+n_2}\,e^{-i\vec{K}_- \cdot \vec{\delta}_1} +
 {a}_{k+n_2}\,e^{-i\vec{K}_-\cdot \vec{\delta}_2 }
 \nonumber\\
 &&+ {a}_{k}\,e^{-i\vec{K}_-\cdot \vec{\delta}_3  })+ U b^*_{k}b_{k}b_{k}  \label{eqn:bRotate2}
 \end{eqnarray}

To summarize, a rotation by $\pi /3 $ which takes  $\vec{K}_+ \rightarrow   \vec{K}_+^\prime$ is identical to the unrotated theory but with $\vec{K}_+ \rightarrow   \vec{K}_- $.  Note that the redefinition of the index $k$ is different for the two equations because old and new primitive cells are related by a rotation.
Rotating by $\pm 2\pi /3$ exchanges A and B sites once more and returns $\Psi$ to its original configuration.
Since $\Psi$ has four components, the effect of making one full rotation of $2\pi $ is that the components acquire a net phase so that   $ \Psi \rightarrow - \Psi$.  This \emph{Berry phase}~\cite{zhangYB2005} endows  $ \Psi$ with the characteristic double-valuedness of a genuine 4-spinor.

However, we must be cautious when discussing chirality and helicity, since we treat (2+1) dimensions and can use only the first two Pauli matrices.  As in the (3+1) theory, one can define a \emph{pseudo-chirality operator} in (2+1) dimensions,  $\gamma^5$, as the product of the other four $\gamma$ matrices.  In the Weyl, or Chiral representation we have
\begin{eqnarray}  \gamma^5 \equiv i \gamma^0 \gamma^1 \gamma^2 \gamma^3 =  \left[
  \begin{array}{ c c }
     {\bf 1} & {\bf 0} \\
       {\bf 0} &  {\bf -1} \\
  \end{array} \right]\,,\end{eqnarray}
where again the boldface indicates a $2\times 2$ submatrix.  This is the natural representation for $\Psi$ in that the NLDE of Eq.~(\ref{eqn:nlde4vec3}) maps into this representation in a natural way:  states of well defined chirality correspond to the upper and lower 2-spinors.  Thus the upper and lower spinors
\begin{equation}
 \Psi_+  \equiv \left[
  \begin{array}{ c  }
      \psi_+ \\
       {\bf 0}
  \end{array} \right]\,,\:\:\:
  \Psi_-  \equiv \left[
  \begin{array}{ c  }
    { \bf  0 } \\
        \psi_-
  \end{array} \right]
  \end{equation}
are eigenfunctions of $ \gamma^5$,
\begin{equation}
\gamma^5  \Psi_+  =  \Psi_+\,, \:\:\:\gamma^5 \Psi_-  = -  \Psi_-\,.
\end{equation}

The question of Poincar\'e covariance of the nonlinear Dirac equation remains.
First, we check coordinate translations.
The wave function is required to transform as
 \begin{eqnarray} \Psi'(r_{\mu} +   d_{\mu}) =  \Psi (r_{\mu})\,.
 \end{eqnarray}
Thus the nonlinear terms which have two factors of $\Psi$ are invariant under translations.
Under spatial rotations we observe that the interaction terms in the NLDE remain unchanged within the context of the full theory.  We include both $\vec{K}_+$ and $\vec{K}_-$ points in the full theory.
As for the case of boosts, we note that for the linear equation the components of the wave function transform in accordance with the transformation of the space-time coordinates in their arguments.  This is exactly canceled by the reciprocal transformations of the partial derivatives.  This is not the case for the nonlinear terms.  One has some matrix product with two factors of the wave function.  Thus the {\it nonlinear} Dirac equation is not invariant under Lorentz boosts.

For any truly fundamental theory of nature it is demanded that the governing equations be invariant with respect to the Poincar\'e group.  Sometimes this requirement is loosened as in the quantization of gauge theories, when fixing the gauge causes relativistic invariance to be non-manifest.  Yet the theory itself certainly remains invariant. Other times the breaking of Poincar\'e covariance implies that we are dealing with an effective theory in which deeper physical processes are at work.  Our case is an example of the latter.  By including on-site interactions we break  relativistic covariance of the linear equation of motion by introducing self-interactions which can be viewed as evidence of ``deeper physics'' in our two-dimensional universe.

\subsection{Hermiticity}

We require the Hamiltonian be Hermitian in order to guarantee that physically measurable quantities (observables) are real.  Thus each term must be independently Hermitian.  In particular, we must show that $N^\dagger = N$.
The proof follows:
\begin{eqnarray}N^\dagger &=&  ( U \Psi^\dagger A_+ \Psi A_+ )^\dagger \nonumber\\
 &=& U A_+^\dagger \Psi^\dagger A_+^\dagger (\Psi^\dagger)^\dagger  \nonumber \\
 &=& U A_+\Psi^\dagger A_+\Psi \nonumber \\
 &=& U\Psi^\dagger A_+\Psi A_+ \nonumber \\
 &=& N \end{eqnarray}
where the last two steps work because $A_+$ is real and symmetric. The nonlinear terms, and therefore the NLDE, are indeed Hermitian.

\subsection{Current Conservation}
\label{ssec:currentConservation}

Current conservation is expected in ordinary QED for a closed system, i.e., an isolated volume of space.
However, most theories of quantum gravity do introduce Lorentz violating terms which bring in charge non-conservation effects.
The conserved current for the linear Dirac equation is
\begin{eqnarray}j^\mu = \bar{\Psi} \gamma^\mu {\Psi} \end{eqnarray}
where $ \bar{\Psi}  \equiv \Psi^\dagger \gamma^0 $.

We check that current  is also conserved for the NLDE.  The 3-divergence of the current
in (2+1) spatial dimensions is
\begin{eqnarray}
 \partial_\mu j^\mu &=& \partial_\mu (\bar{\Psi} \gamma^\mu \Psi)    \nonumber \\
 &=& \partial_\mu ({\Psi}^\dagger \gamma^0 \gamma^\mu \Psi)  \nonumber\\
 &=& -   (\partial_\mu {\Psi}^\dagger) \gamma^\mu \gamma^0 \Psi  +   {\Psi}^\dagger   \gamma^0 \gamma^\mu\partial_\mu \Psi  \label{eqn:current1}\end{eqnarray}
where $\mu\in\{0,1,2\}$ and we have used the chain rule and the anti-commuting properties of the Dirac matrices.
Taking the adjoint of the NLDE in four-component form, Eq.~(\ref{eqn:nlde4vec3}), yields
 \begin{equation}
 {( i \gamma^\mu \partial_\mu \Psi)}^\dagger  = {( - \gamma^0 N \Psi)}^\dagger\,,
 \end{equation}
which implies
\begin{equation}
-  \partial_\mu \Psi^\dagger \gamma^\mu = -i \Psi^\dagger N^\dagger \gamma^0\,.
\end{equation}
Then the (2+1)-divergence of the current is
 \begin{eqnarray}
 \partial_\mu j^\mu  &=& -  i \Psi^\dagger N^\dagger \gamma^0 \gamma^0 \Psi +  \Psi^\dagger \gamma^0 (i \gamma^0 N \Psi) \\
 &=& -  i \Psi^\dagger (N^\dagger  -  N) \Psi = 0
\end{eqnarray}
Thus current is conserved.  This is in fact the statement that hermiticity implies current conservation.

\subsection{Chiral Current}

The conserved chiral current for the linear Dirac equation is
 \begin{equation} j_5^\mu \equiv \bar{\Psi} \gamma^\mu \gamma_5 \Psi\,. \end{equation}
Then the (2+1)-divergence of the chiral current is
 \begin{equation} \partial_\mu  j_5^\mu =  \partial_\mu  ( \bar{\Psi} \gamma^\mu \gamma_5 \Psi)\,.
 \end{equation}
Following a similar route as in Sec.~\ref{ssec:currentConservation}, we find
\begin{equation}
 \partial_\mu  j_5^\mu = -  i \Psi^\dagger (N \gamma_5  + \gamma_5  N) \Psi
\end{equation}
where we have used the hermiticity of $N$.  This means that in order for chiral current to be conserved we must have
 \begin{eqnarray} \{N , \gamma_5\}  =  0 \end{eqnarray}
where the curly braces signify the anti-commutator.  The anti-commutator is
 \begin{equation}
 \{N_+ , \gamma_5\} = U \Psi^\dagger {\it A_+} \Psi {\it A_+} \gamma_5 +   \gamma_5  U \Psi^\dagger {\it A_+} \Psi {\it A_+}\,.
 \end{equation}
Writing this out in explicit matrix form we find that
 \begin{eqnarray}
 \{N_+ , \gamma_5 \} &=& + 2 N\,,\\
 \{N_- , \gamma_5\} &=& - 2 N\,.
 \end{eqnarray}
Therefore chiral current is not conserved. This fact is reminiscent of the anomalous non-conservation of chiral current in the case of some field theories upon quantization, which is mediated by instantons.  It is interesting to consider that our nonlinear terms might be treated as quantum-induced nonlinearities.

\subsection{Universality}

Universality refers to the invariance of a theory under rescaling of the solution. For the case of the linear Dirac equation,  $\Psi \rightarrow \lambda \Psi$ leaves the theory unchanged.  Because the nonlinear term in the NLDE contains one factor of $\Psi$ and one factor of its adjoint, it scales as $\lambda^2$, thus breaking the invariance of the nonlinear theory.  For a treatment of NLDEs which are universal in this sense, but not relevant to the present solid state system, see Ref.~\cite{ngWK2007}.

\subsection{Discrete Symmetries}

\subsubsection{Parity}

In the case of the standard massless Dirac equation, invariance under the parity operation $\hat{P}$ requires $\Psi$ to transform as
 \begin{equation}
 \Psi \rightarrow  \Psi^\prime = {\it \hat{ P}}  \Psi= {\gamma^0} \Psi \,,
 \label{eqn:parityCondition}
 \end{equation}
where, in contrast to the Chiral representation presented in Eq.~(\ref{eqn:diracMatrices}), we take $\gamma^0$ in this section to be in the Dirac representation, 
\begin{eqnarray} \gamma^0 = \left[
  \begin{array}{ c c }
      {\bf1} &  {\bf0}  \\
      {\bf 0} & - {\bf1}
       \end{array} \right]   \end{eqnarray}

We proceed to determine if the NLDE remains invariant under the transformation of Eq.~(\ref{eqn:parityCondition}).  The parity operator acting on the honeycomb lattice inverts both coordinate axes, $\vec{\nabla} \rightarrow - \vec{\nabla}$, and thus exchanges A and B sites while also exchanging $\vec{K}$-point indices.  The transformed \emph{linear} equations ($U=0$) are
 \begin{eqnarray}i {\bf \gamma^0} \left[
  \begin{array}{ c }
     \dot{\Psi}_{-} \\
     \dot{\Psi}_{+}
   \end{array} \right]
  + ic_s  \left[
  \begin{array}{ c c c}
      -\bf{\vec{\sigma} \cdot \vec{\nabla}} &  \bf{ 0} \\
       \bf{0}   &  \vec{\sigma} \cdot \vec{\nabla}
  \end{array} \right] \gamma^0  \left[
  \begin{array}{ c  }
       \Psi_{-} \\
       \Psi_{+}
  \end{array} \right]    = 0 \,.\label{eqn:linearParity1}\end{eqnarray}
Interchanging upper and lower spinors we obtain the equivalent form
  \begin{eqnarray} i \left[
  \begin{array}{ c }
     \dot{\Psi}_{+} \\
     \dot{\Psi}_{-}
     \end{array} \right]
  + ic_s  \left[
  \begin{array}{ c c c}
      \bf{\vec{\sigma} \cdot \vec{\nabla}} &  \bf{ 0} \\
       \bf{0}   &  -\vec{\sigma} \cdot \vec{\nabla}
  \end{array} \right]  \left[
  \begin{array}{ c  }
      \Psi_{+} \\
      \Psi_{-}
  \end{array} \right]    = 0\,.\label{eqn:linearParity2}\end{eqnarray}
Thus the linear equations are invariant under parity.

Next we check the nonlinear term in the NLDE.
With the transformed nonlinear term added to Eq.~(\ref{eqn:linearParity1}) we obtain
\begin{eqnarray}i \left[
  \begin{array}{ c }
     \dot{\Psi}_{-} \\
   - \dot{\Psi}_{+}
  \end{array} \right]
&&   + ic_s  \left[
  \begin{array}{ c c c}
      -\bf{\vec{\sigma} \cdot \vec{\nabla}} &  \bf{ 0} \\
       \bf{0}   &  \vec{\sigma} \cdot \vec{\nabla}
  \end{array} \right]  \left[
  \begin{array}{ c  }
      \Psi_{-} \\
     - \Psi_{+}
     \end{array} \right]  \nonumber\\
     &&- U  \left[
  \begin{array}{ c }
     N_-\\
     - N_+\\
  \end{array} \right] = 0 \,, \end{eqnarray}
which after similar steps as above gives us back the NLDE.  Therefore parity is a symmetry of the nonlinear Dirac equation.  We note that parity is conserved in standard QED and also in the theory of the strong interactions, but violated by weak interactions.  Therefore, in this aspect, the NLDE is consistent with QED.

\subsubsection{Charge Conjugation}

In order to maintain symmetry under charge conjugation one requires the nonlinear term to transform as
   \begin{eqnarray}
   {(\gamma^0 N)}^\prime  =  {\it \hat{C} }  \gamma^0     {N}^*   {\it \hat{C}^{-1}}
   \end{eqnarray}
where in the Chiral representation the charge conjugation operator is
  \begin{eqnarray} {\it \hat{C} } =  \gamma^1   \nonumber  \end{eqnarray}
and the wavefunction transforms in the standard way:
  \begin{eqnarray}
  \Psi \rightarrow \gamma^1 \gamma^0 {\Psi^{\dagger}}^T\,.
  \end{eqnarray}
We determine if the NLDE maintains this symmetry.  Let $F$ be one of the nonlinear terms.  Then
   \begin{eqnarray}  {(\gamma^0 N)}^\prime  &=& {(\gamma^1 \gamma^0 {\Psi^{\dagger}}^T)}^\dagger A_+ (\gamma^1 \gamma^0 {\Psi^{\dagger}}^T)  A_+ \nonumber \\
  &=& \Psi^T \gamma^0 \gamma^1 A_+ \gamma^1 \gamma^0 \Psi^* A_+   \end{eqnarray}
Also, we have
 \begin{eqnarray} {\it \hat{C} }  \gamma^0     {N}^*   {\it \hat{C}^{-1}}  &=&  \gamma^1 \gamma^0{( \Psi^\dagger A_+ \Psi A_+)}^* \gamma^1 \nonumber\\
 &=& \gamma^1 \gamma^0  \Psi^T A_+ \Psi^*  A_+ \gamma^1
\end{eqnarray}
Thus the NLDE breaks charge conjugation symmetry.

\subsubsection{Time Reversal}

The usual time reversal, $t \rightarrow - t$, requires that $\Psi$  transform as
 \begin{equation}
\Psi \rightarrow \Psi' = \hat{\Theta} \Psi =   i{ \gamma^1}{ \gamma^3}  { \Psi}\,,
 \end{equation}
where $\hat{\Theta}$ is the time-reversal operator and
 \begin{eqnarray}i{ \gamma^1} {\gamma^3} = \left[
  \begin{array}{ c c }
       -  \sigma_y &  {\bf 0} \\
      {\bf 0} &  -  \sigma_y  \\
  \end{array} \right]\,,\end{eqnarray}
 In our theory the intrinsic effect of time reversal is to change the direction of momentum so that
$\vec{K}$ points are switched without exchanging A and B indices. Combining these effects, we determine whether or not the linear part of the NLDE, i.e., the linear Dirac equation,  remains invariant:
\begin{eqnarray}&&-i \frac{\partial}{\partial t}  i{\bf \gamma^1}{\bf \gamma^3}   \left[
  \begin{array}{ c }
     \Psi_{-} \\
     \Psi_{+}
  \end{array} \right]\nonumber\\
&&   + ic_s   \left[
  \begin{array}{ c c c}
      \bf{\vec{\sigma} \cdot \vec{\nabla}} &  \bf{ 0} \\
       \bf{0}   &  -\vec{\sigma} \cdot \vec{\nabla}
  \end{array} \right]  i{\bf \gamma^1}{\bf \gamma^3} \left[
  \begin{array}{ c  }
      \Psi_{-} \\
      \Psi_{+}
  \end{array} \right]   = 0 \end{eqnarray}
Then
 \begin{eqnarray} -  i \frac{\partial}{\partial t}    \left[
  \begin{array}{ c }
     \Psi_{+} \\
     \Psi_{-}
  \end{array} \right]
  + ic_s   \left[
  \begin{array}{ c c c}
      \bf{\vec{\sigma} \cdot \vec{\nabla}} &  \bf{ 0} \\
       \bf{0}   &  -\vec{\sigma} \cdot \vec{\nabla}
  \end{array} \right]  \left[
  \begin{array}{ c }
     \Psi_{+} \\
     \Psi_{-}
  \end{array}  \right] = 0 \end{eqnarray}
The appearance of the negative sign in front of the time derivative indicates that these are the equations satisfied by the negative energy solutions, or holes.  Thus time reversal takes the equations describing electrons into holes and those for holes into electrons but keeps the overall theory invariant.

We proceed to consider the nonlinear term.  With the transformed nonlinear term we obtain
  \begin{eqnarray}
  &&-  i \frac{\partial}{\partial t}    \left[
  \begin{array}{ c }
     \Psi_{+} \\
     \Psi_{-}
  \end{array} \right]
  +ic_s   \left[
  \begin{array}{ c c c}
      \bf{\vec{\sigma} \cdot \vec{\nabla}} &  \bf{ 0} \\
       \bf{0}   &  -\vec{\sigma} \cdot \vec{\nabla}
  \end{array} \right]  \left[
  \begin{array}{ c }
     \Psi_{+} \\
     \Psi_{-}
  \end{array}  \right]\nonumber\\
  &&-  i U  \left[
  \begin{array}{ c }
     N_+ \\
     N_-  \\
     \end{array} \right]    = 0  \,.\label{eqn:timeReversal}\end{eqnarray}
 In Eq.~(\ref{eqn:timeReversal}) the interaction term has acquired a factor of $i$.  Therefore the NLDE is not invariant under time reversal.  Thus, as one also observes in the Standard Model of particle physics, CP and T are not conserved independently but CPT is conserved.

\section{Discussion and Conclusions}
\label{sec:4}

The nonlinear Dirac equation we have presented introduces a completely new class of nonlinear phenomena in Bose-Einstein condensates.  Although our work is related to graphene, in that the BEC is taken to be trapped on a honeycomb lattice, we have switched bosons for fermions.  The form of the nonlinearity is then a natural physical result of binary interactions between bosons.  In fact, it is a spinor generalization of the kind of nonlinearity one finds in the nonlinear Schrodinger equation, i.e., proportional to the local condensate density.  The same equation will occur for light subject to a Kerr nonlinearity when propagating through a photonic crystal with a honeycomb lattice structure~\cite{akozbekN1998}.

We showed that the NLDE breaks Poincar\'e covariance, and therefore the principle of relativity.  This suggests that small nonlinearities of this form could be looked for by such symmetry breaking in a variety of systems where Dirac or Dirac-like equations apply.  For instance, even for fermions there is a small mean field effect.  We could just as easily have considered ultra-cold fermions on an optical lattice.  This would appear at first sight to be the exact analog of graphene; however, our work shows that Poincar\'e covariance will be broken by mean field effects, even if on a small level.  Indeed, for graphene one should expect similar effects due to Coulomb interactions. The latter nonlinearity can be expected to be non-local due to the power law behavior of $1/r$ for the Coulomb potential, just as dipole-dipole interactions between ultra-cold atoms even without an optical lattice lead to a non-local nonlinear Schrodinger equation in a 3D continuum (in 2D dipole-dipole interactions, which have a power law of $1/r^3$, are local).  Thus, in graphene, we make the conjecture that there is a non-local nonlinear correction term to the massless Dirac equation which breaks Poincar\'e covariance.  Similar corrections can be looked for in quantum electrodynamics under the proper circumstances~\cite{ferrandoA2007}.

All generalizations of the scalar nonlinear Schrodinger equation relevant to BECs are candidates for generalized nonlinear Dirac equations.  For example, pseudo-spin structure leads to a vector NLSE.  One can therefore anticipate a vector NLDE as well.

We have thoroughly explored both continuous and discrete symmetries of the NLDE.  In particular, we showed that pseudo-chiral current is not conserved, the NLDE is not covariant under Lorentz boosts, and it breaks charge conjugation  as well as time reversal symmetry.  On the other hand, the NLDE is hermitian, local, conserves current, and is symmetric under
parity.  The NLDE maintains CPT symmetry, just as one finds in the Standard Model.

In a future work we will treat soliton and vortex solutions of the NLDE, as a first step towards a complete classification of nonlinear relativistic phenomena in BECs.

We acknowledge useful discussions with Alex Flournoy and Mark Lusk.
This work was supported by the National Science
Foundation under Grant PHY-0547845 as part of the NSF CAREER program.

% \appendix

% \section{}
% \label{}

%\begin{thebibliography}{00}


% notes:

% \bibitem{label2}
% If there is a note, it should come last:
% \bibitem{label3} \note
% \end{subbibitems}

%\bibitem{}

%\end{thebibliography}

%\bibliographystyle{elsart-num}
%\bibliography{refs}

\end{document}